\documentclass[3p,times,procedia]{elsarticle}
\usepackage{nupha_ecrc}

\volume{00}

\firstpage{1}

\journalname{Nuclear Physics A}

\runauth{}

\jid{nupha}

\jnltitlelogo{Nuclear Physics A}

\usepackage{amssymb}

\usepackage{lineno}

\usepackage[figuresright]{rotating}

\usepackage{upgreek}
\usepackage{amsmath}
\usepackage{amssymb}

\newcommand{\sNN }{\sqrt{s_{\text{\textrm{NN}}}} }           
\newcommand{\pT}{$p_{T}$}
\newcommand{\AuAu}{Au+Au }
\newcommand{\pp}{$\textrm{p+p}$ }

\begin{document}

\begin{frontmatter}



\dochead{}

\title{ $\Lambda_{\textup{c}}$ Production in $\textup{Au+Au}$ Collisions at $\sqrt{s_{\textup{NN}}}$ = 200 GeV measured by the STAR experiment}


\author{Guannan Xie (for the STAR\fnref{col1}  Collaboration)}
\fntext[col1] {A list of members of the STAR Collaboration and acknowledgments can be found at the end of this issue.}

\address{
University of Science and Technology of China, Hefei, 230026, China

Lawrence Berkeley National Laboratory, Berkeley, CA 94720, USA
}

\begin{abstract}

  At RHIC, enhancements in the baryon-to-meson ratio for light hadrons and hadrons containing strange quarks have been observed in central heavy-ion collisions compared to those in \pp and peripheral heavy-ion collisions in the intermediate transverse momentum (\pT) range (2 $<$ \pT $\ <$ 6 GeV/$c$). This can be explained by the hadronization mechanism involving multi-parton coalescence. $\Lambda_{\textup{c}}$ is the lightest charmed baryon with mass close to that of $D^0$ meson, and has an extremely short life time ($c\tau\sim60 \ \mu\textrm{m}$). Different models predict different magnitudes of enhancement in the $\Lambda_{\textup{c}}$/$D^0$ ratio depending on the degree to which charm quarks are thermalized in the medium and how the coalescence mechanism is implemented.

  In these proceedings, we report the first measurement of $\Lambda_{\textup{c}}$ production in heavy-ion collisions using the Heavy Flavor Tracker at STAR. The invariant yield of $\Lambda_{\textup{c}}$ for 3 $<$ \pT $\ <$ 6 GeV/$c$ is measured in 10-60\% central Au+Au collisions at $\sNN$ = 200 GeV. The $\Lambda_{\textup{c}}$/$D^0$ ratio is compared to different model calculations, and the physics implications are discussed.

\end{abstract}


\end{frontmatter}


\section{Introduction}
\label{Introduction}

Charm quarks are dominantly produced at the early stages of heavy-ion collisions through hard scattering processes at RHIC energies since their masses are significantly larger than the temperature of the quark-gluon plasma (QGP) created in such collisions making the thermal production improbable. Therefore, they experience the entire evolution of the system, and provide unique information on the properties of the hot and dense strongly-coupled QGP.

\begin{figure}[htbp]
\hspace{+2.2cm}
\begin{minipage}[b]{0.75\linewidth}
\begin{center}
\includegraphics[width=\textwidth]{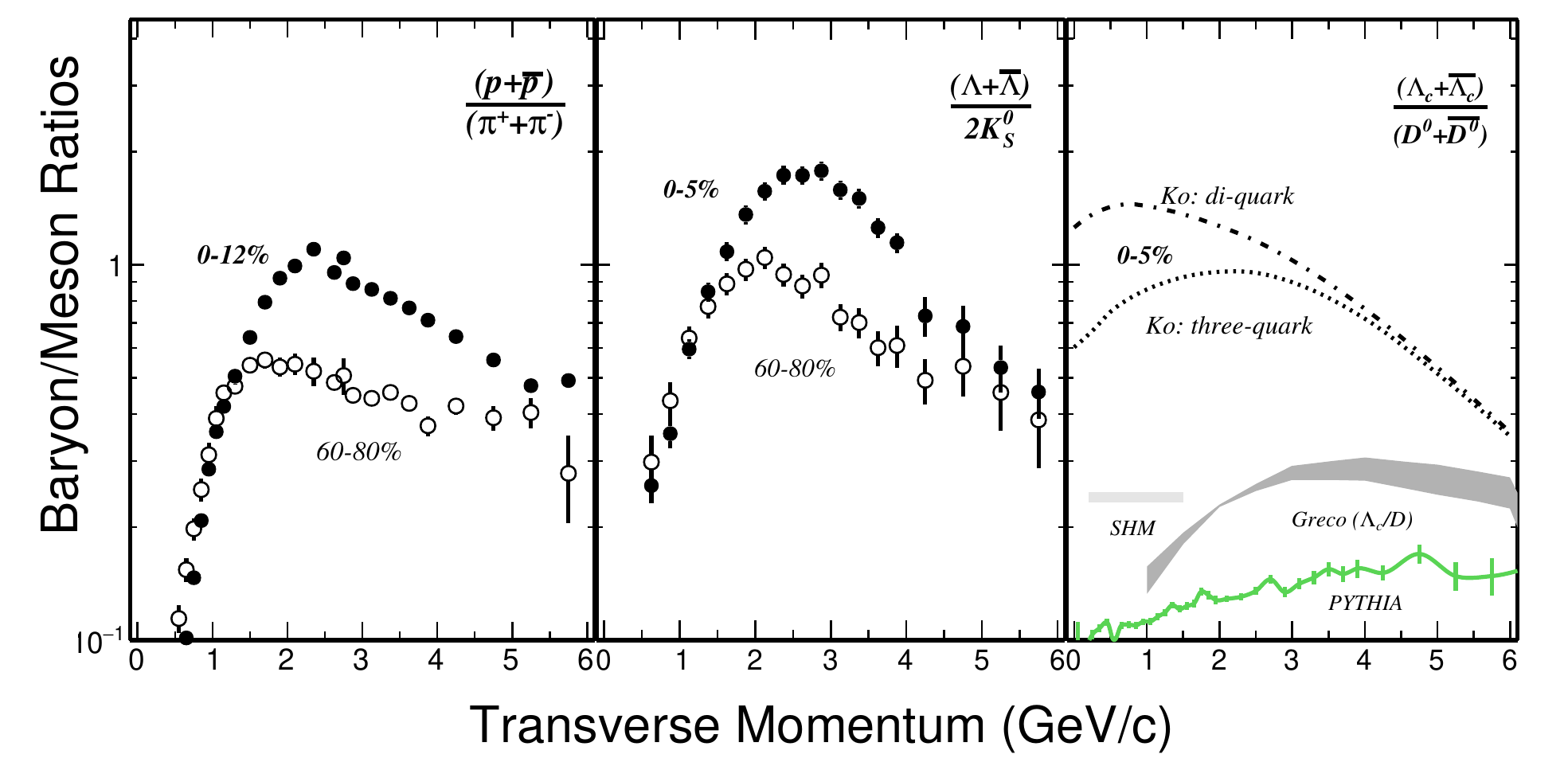}
\end{center}
\end{minipage}
\caption{Baryon-to-meson ratio for $p/\pi$ (left panel) and $\Lambda$/$K_s$ (middle panel) measured in \AuAu collisions by the STAR experiment \cite{STAR2006, Adams2006, Abelev2009}. The right panel shows several model predictions for $\Lambda_{\textup{c}}$/$D^0$ ratio \cite{Andronic, Kuznetsova2007, Andronic2008, Oh2009, Lee2008, Ghosh2014}.}
\label{fig:BtoMratio1}
\end{figure}

Measurements on the baryon-to-meson ratios for light hadrons and hadrons containing strange quarks have shown significant enhancements in central heavy-ion collisions compared to those in \pp and peripheral heavy-ion collisions in the intermediate transverse momentum (\pT) range as illustrated in Fig.~\ref{fig:BtoMratio1} \cite{STAR2006, Adams2006, Abelev2009}. The enhancement can be explained by the coalescence hadronization through recombination of constituent quarks \cite{Greco2003a, Greco2003}. Charm quarks may hardronize through coalescence too, which might also manifest itself through an enhancement in the baryon-to-meson ratio for charm hadrons in central heavy-ion collisions. Theoretical calculations for such an enhancement would be sensitive to how the coalescence mechanism is implemented and the degree to which charm quarks are thermalized in the medium. Experimental measurements are crucial for distinguishing these models and shed lights on the charm quark hadronization in the hot and dense medium.

\section{Experiment and Analysis}
\label{experiment}

The STAR experiment at the RHIC is a large-acceptance detector covering full azimuth and pseudorapidity of $|\eta| < 1$. Data were taken by the STAR experiment with the Heavy Flavor Tracker (HFT) installed in 2014. The HFT is a high resolution silicon detector providing a track pointing resolution of less than 40 $\mu\textrm{m}$ for kaons with \pT $\ $= 1 GeV/$c$. It has successfully completed its scientific mission during the years 2014 - 2016, and been taken out of the STAR experiment since then.

In total, about 900M minimum bias \AuAu events are used for this analysis. These events are selected to contain primary vertices (PV) within 6 cm to the center of the STAR detector along the beam direction to ensure uniform HFT acceptance. Both $\Lambda^{+}_{\textup{c}}$ and $\Lambda^{-}_{\textup{c}}$ (in the following, $\Lambda_{\textup{c}}$ represents the sum of the two charge conjugates) are reconstructed through the hadronic decay channel ($\Lambda_{\textup{c}}$ $\rightarrow$ pK$\pi$), with a branching ratio of $\sim6.35\%$ and a proper decay length of $c\tau\sim60$ $\mu\textrm{m}$. Kaons, pions and protons are identified via the ionization energy loss (dE/dx) measured by the Time Projection Chamber (TPC) and the time of flight measured by the Time-Of-Flight (TOF) detector \cite{Beringer2012, Ackermann2003}.

$\Lambda_{\textup{c}}$ decay vertices are reconstructed with selected kaon, pion and proton tracks. The middle points on the distance of closest approach (DCA) are calculated for each pair of daughter candidate tracks. The centroid of the three middle points for the three pairs is used as the decay vertex. With the HFT, the following 6 geometrical variables are used to select $\Lambda_{\textup{c}}$ candidates and suppress the combinatorial background which is dominated by tracks produced at the PV: decay length (the distance between the decay vertex and the PV), the maximum DCA between three daughters, cos($\theta$) where $\theta$ is the back pointing angle between the $\Lambda_{\textup{c}}$ momentum and the decay vertex direction w.r.t. the PV, DCA between the pion track and the PV, DCA between the kaon track and the PV and DCA between the proton track and the PV. The cut values on these variables are optimized using the Toolkit for Multivariate Data Analysis (TMVA) package to achieve the best significance for the $p_T$ range of 3 to 6 GeV/$c$. Additionally, a maximum DCA cut between the $\Lambda_{\textup{c}}$ candidate and the PV is applied to make sure the $\Lambda_{\textup{c}}$ candidate is produced at the PV.

\begin{figure}[htbp]
\hspace{+0.5cm}
\begin{minipage}[b]{0.45\linewidth}
\begin{center}
\includegraphics[width=\textwidth]{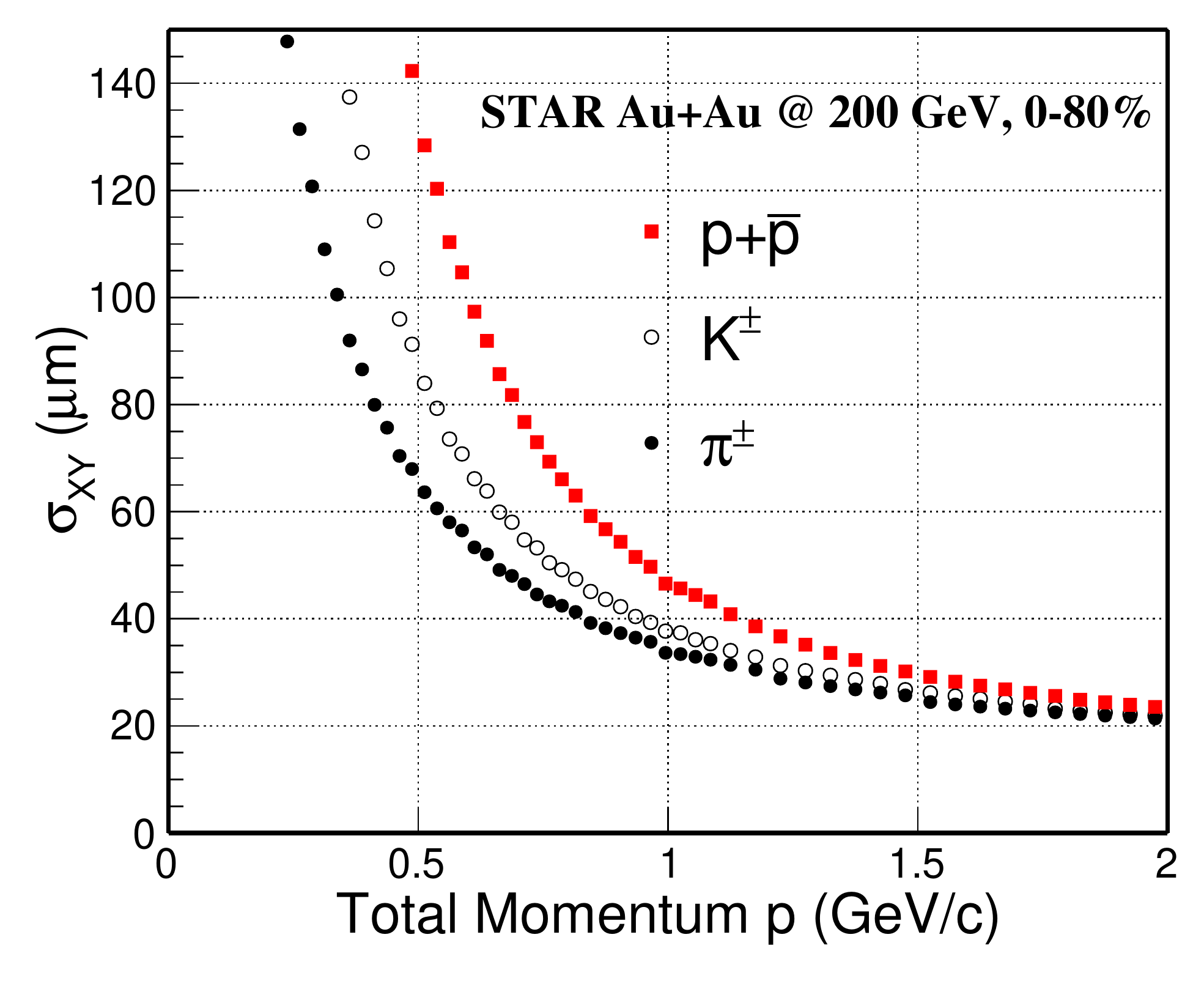}
\end{center}
\end{minipage}
\hspace{+0.0cm}
\begin{minipage}[b]{0.45\linewidth}
\begin{center}
\includegraphics[width=\textwidth]{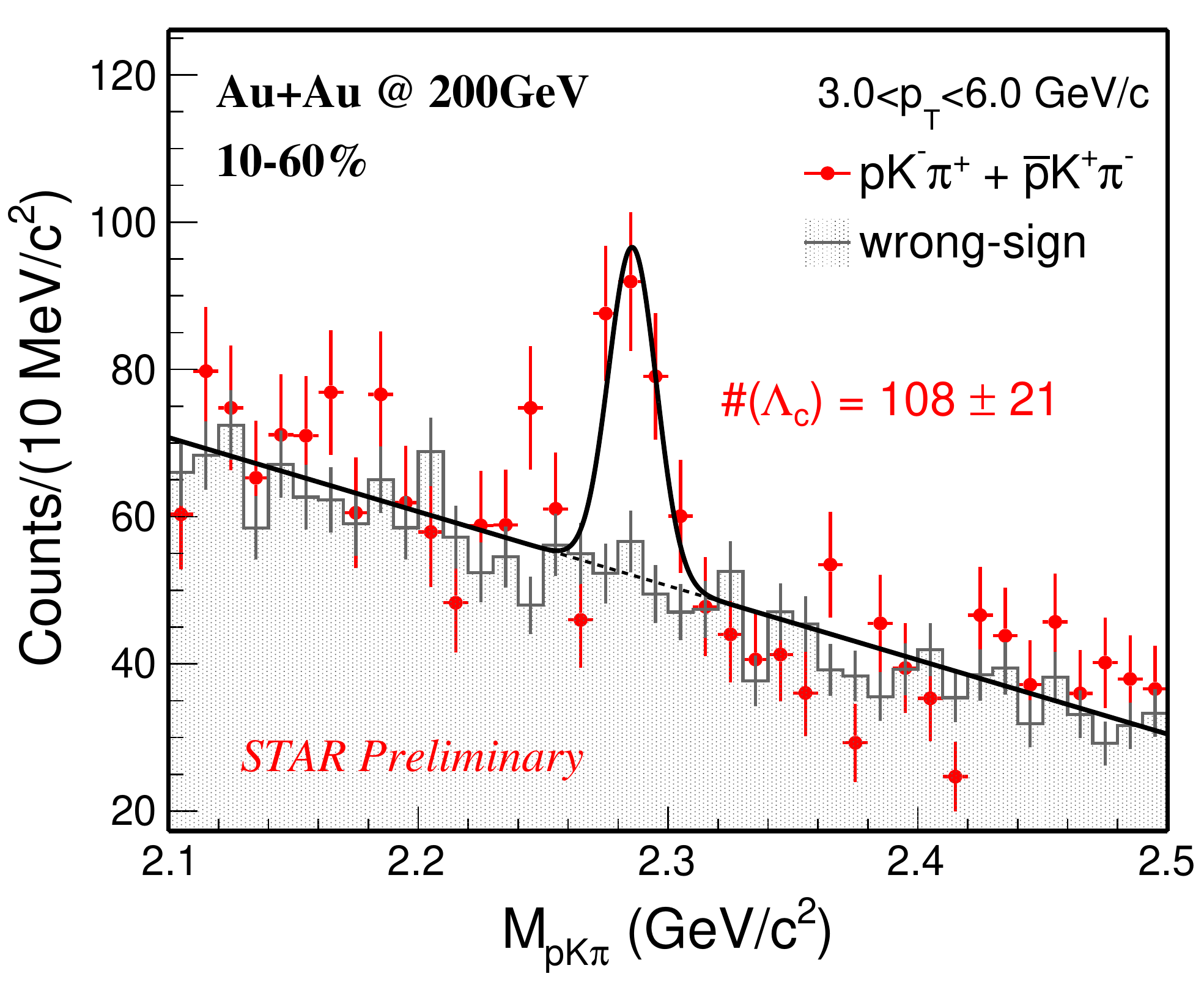}
\end{center}
\end{minipage}
\caption{ (Left) $\textup{DCA}_{\textup{XY}}$ resolution with the STAR HFT achieved in 2014 for pions (solid circles), kaons (open circles) and protons (red solid squares). (Right) Invariant mass spectra of K$\pi$p pairs for 3 $<$ \pT $\ <$ 6 GeV/$c$. The red solid data points are the $\Lambda_{\textup{c}}$ signals reconstructed with right-sign combinations. The grey histogram depicts the wrong-sign background distribution (scaled by 1/3).}
\label{fig:DcaXyLcSignal}
\end{figure}

Fig.~\ref{fig:DcaXyLcSignal}, left panel, shows the DCA resolution in the transverse plane achieved in 2014 data for pions, kaons and protons, which exceeds the design goal of 55 $\mu\textrm{m}$ resolution for kaons at \pT $\ $= 750 MeV/$c$. Fig.~\ref{fig:DcaXyLcSignal}, right panel, shows the invariant mass spectra of pK$\pi$ triplets for 3 $<$ \pT $\ <$ 6 GeV/$c$ in 10-60\% central Au+Au collisions. Red markers represent the right sign combinations for $\Lambda_{\textup{c}}$ candidates while the grey histogram depicts the wrong-sign combinations scaled by 1/3.

The reconstruction efficiency is calculated in two parts. The TPC tracking efficiency and acceptance are calculated using the standard embedding procedure. On the other hand, the HFT tracking efficiency, acceptance and the geometric cut efficiencies are estimated from a fast simulation based on the HFT to TPC matching ratios and the track pointing resolutions extracted directly from data.

\section{Physics Results and Discussion}
\label{results}

\begin{figure}[htbp]
\hspace{+3.0cm}
\begin{minipage}[b]{0.48\linewidth}
\begin{center}
\includegraphics[width=\textwidth]{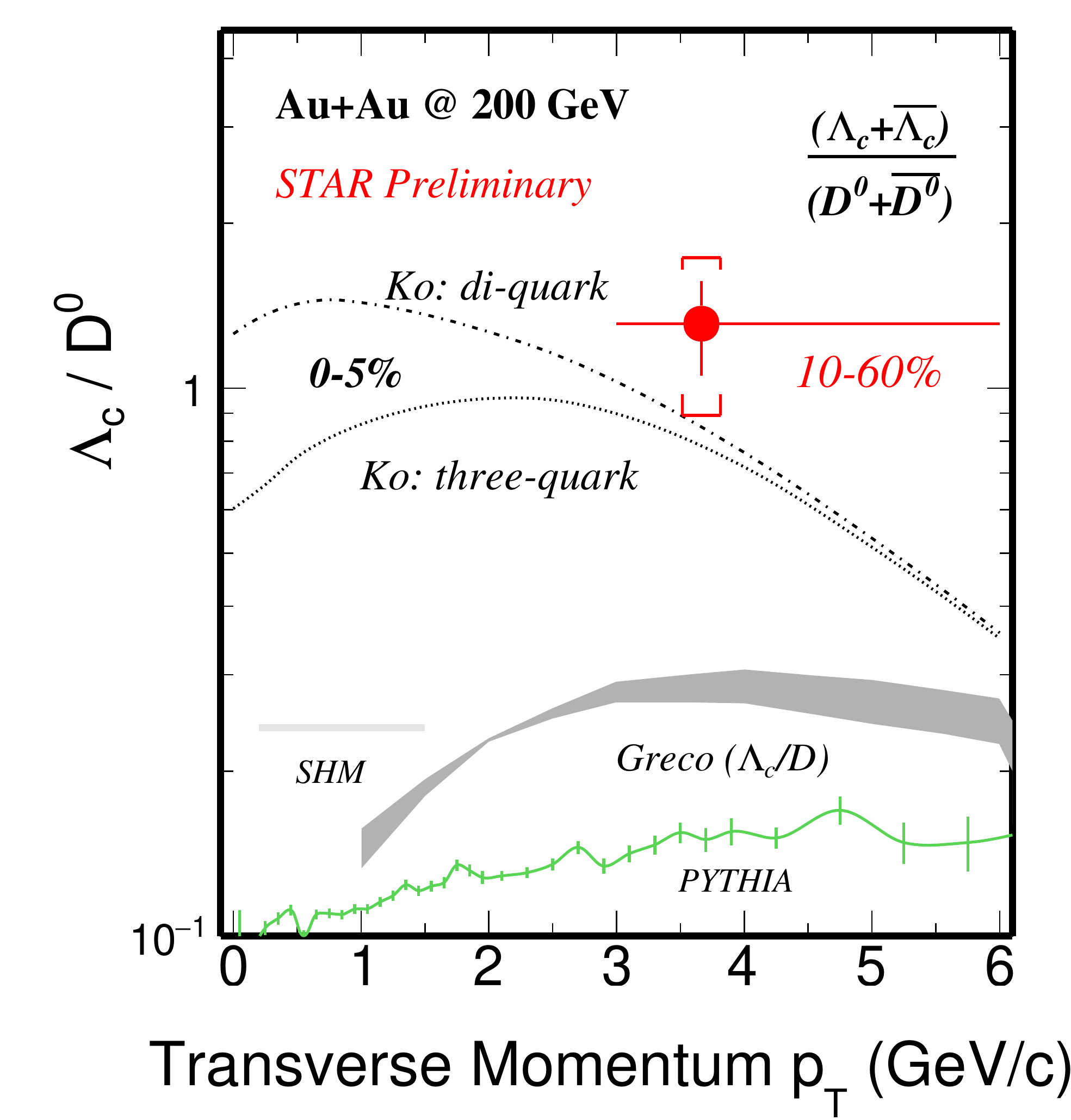}
\end{center}
\end{minipage}
\caption{Measured $\Lambda_{\textup{c}}$/$D^0$ ratio for 3 $<$ \pT $\ <$ 6 GeV/$c$ in 10-60\% central \AuAu collisions at $\sNN$ = 200 GeV compared to model predictions from PYTHIA \cite{Sjostrand2006}, statistical hadronization model (SHM) \cite{Andronic, Kuznetsova2007, Andronic2008} and coalescence models from Ko and Greco \cite{Oh2009, Lee2008, Ghosh2014}.} 
\label{fig:BtoMratio2}
\end{figure}

After correcting for the acceptance and efficiency, one can obtain the $\Lambda_{\textup{c}}$ yield and calculate the $\Lambda_{\textup{c}}$/$D^0$ ratio. Fig.~\ref{fig:BtoMratio2} shows the first measurement of the $\Lambda_{\textup{c}}$/$D^0$ ratio for 3 $<$ \pT $\ <$ 6 GeV/$c$ in 10-60\% central \AuAu collisions at $\sNN$ = 200 GeV along with model predictions. The measured $\Lambda_{\textup{c}}$/$D^0$ ratio in \AuAu collisions is significantly enhanced compared to the PYTHIA prediction for \pp collisions \cite{Sjostrand2006}. The enhancement is also larger than the statistical hadronization model (SHM) predictions \cite{Andronic, Kuznetsova2007, Andronic2008}. In Ko's model, thermalized charm quarks are used for recombination and the predicted $\Lambda_{\textup{c}}$/$D^0$ ratio is comparable to our measurement despite that the calculation is done for the 0-5\% centrality bin. However, one needs measurements at low \pT $\ $to further differentiate between three-quark and di-quark recombination scenarios \cite{Oh2009, Lee2008}. In Greco's model, charm quarks diffuse in the QGP medium and then recombine with light quarks to form charm hadrons \cite{Ghosh2014}. The calculated ratio is w.r.t the total charm meson yield including $D^0$, $D^{\pm}$ and $D_s^{\pm}$. One may expect a factor of 1.5 (\pp baseline) or larger (if $D_s$ is enhanced) increase to be compared to our $\Lambda_{\textup{c}}$/$D^0$ ratio. The observed $\Lambda_{\textup{c}}$/$D^0$ ratio is comparable to the baryon-to-meson ratios for light hadrons and strangeness hadrons shown in Fig.~\ref{fig:BtoMratio1}.


\section{Summary and Outlook}
\label{summary}

We report the first measurement of $\Lambda_{\textup{c}}$ production in heavy-ion collisions by the STAR experiment. The measured $\Lambda_{\textup{c}}$/$D^0$ ratio shows a significant enhancement for 3 $<$ \pT $\ <$ 6 GeV/$c$ in 10-60\% central \AuAu collisions at $\sNN$ = 200 GeV compared to the PYTHIA calculation. It is interesting to note that the measured $\Lambda_{\textup{c}}$/$D^0$ ratio is comparable to the measured baryon-to-meson ratios for light and strangeness hadrons. A coalescence model calculation using thermalized charm quarks is consistent with the measurement. In 2016, STAR has collected 2 billion minimum-bias \AuAu events with the HFT which will allow for more precise measurements in the near future.

\section*{Acknowledgement}
From USTC, the author is supported in part by the NSFC under Grant No.s 11375172, 11375184 and 11675168, and MoST of China under No. 2014CB845400.



\bibliographystyle{elsarticle-num}
\bibliography{GuannanX_bib}



\end{document}